\documentclass[%
 reprint,
 amsmath,amssymb,
 aps,
]{revtex4-2}

\usepackage[dvipsnames]{xcolor}
\usepackage{graphicx}
\usepackage{dcolumn}
\usepackage{bm}
\usepackage{cleveref}

\begin{document}


\title{Physics-based inverse modeling of battery degradation with Bayesian methods} 

\author{Micha C. J.~Philipp}
 \affiliation{Institute for Engineering Thermodynamics, German Aerospace Center (DLR), Wilhelm-Runge-Straße 10, 89081 Ulm, Germany}
 \affiliation{Helmholtz Institute Ulm (HIU), Helmholtzstraße 11, 89081 Ulm, Germany}
 \affiliation{Faculty of Natural Sciences, Ulm University, Albert-Einstein-Allee 11, 89081 Ulm, Germany}
\author{Yannick Kuhn}
 \affiliation{Institute for Engineering Thermodynamics, German Aerospace Center (DLR), Wilhelm-Runge-Straße 10, 89081 Ulm, Germany}
 \affiliation{Helmholtz Institute Ulm (HIU), Helmholtzstraße 11, 89081 Ulm, Germany}
 \affiliation{Faculty of Natural Sciences, Ulm University, Albert-Einstein-Allee 11, 89081 Ulm, Germany}
\author{Arnulf Latz}
 \affiliation{Institute for Engineering Thermodynamics, German Aerospace Center (DLR), Wilhelm-Runge-Straße 10, 89081 Ulm, Germany}
 \affiliation{Helmholtz Institute Ulm (HIU), Helmholtzstraße 11, 89081 Ulm, Germany}
 \affiliation{Faculty of Natural Sciences, Ulm University, Albert-Einstein-Allee 11, 89081 Ulm, Germany}
\author{Birger Horstmann}
 \email{birger.horstmann@dlr.de}
 \affiliation{Institute for Engineering Thermodynamics, German Aerospace Center (DLR), Wilhelm-Runge-Straße 10, 89081 Ulm, Germany}
 \affiliation{Helmholtz Institute Ulm (HIU), Helmholtzstraße 11, 89081 Ulm, Germany}
 \affiliation{Faculty of Natural Sciences, Ulm University, Albert-Einstein-Allee 11, 89081 Ulm, Germany}

\date{\today}

\begin{abstract}
To further improve Lithium-ion batteries (LiBs), a profound understanding of complex battery processes is crucial. Physical models offer understanding but are difficult to validate and parameterize. Therefore, automated machine-learning methods (ML) are necessary to evaluate models with experimental data. Bayesian methods, e.g., \emph{Bayesian optimization for likelihood-free inference} (EP-BOLFI), stand out as they capture uncertainties in models and data while granting meaningful parameterization. An important topic is prolonging battery lifetime, which is limited by degradation, such as the solid-electrolyte interphase (SEI) growth. As a case study, we apply EP-BOLFI to parametrize SEI growth models with synthetic and real degradation data. EP-BOLFI allows for incorporating human expertise in the form of suitable feature selection, which improves the parametrization. We show that even under impeded conditions, we achieve correct parameterization with reasonable uncertainty quantification, needing less computational effort than standard Markov chain Monte Carlo methods. Additionally, the physically reliable summary statistics show if parameters are strongly correlated and not unambiguously identifiable. Further, we investigate \emph{Bayesian alternately subsampled quadrature} (BASQ), which calculates model probabilities, to confirm electron diffusion as the best theoretical model to describe SEI growth during battery storage.

\begin{description}
\item[keywords]
machine learning --- bayesian methods --- inverse modeling --- model selection --- lithium-ion batteries --- battery degradation --- SEI formation --- parameterization --- uncertainty quantification
\end{description}
\end{abstract}

\maketitle



\section{Introduction}\label{sec:Introduction}

Lithium-ion batteries (LiBs) are of central importance for the transition to renewable energy sources, especially for the electrification of the transport sector. To meet the high requirements of the transport sector, such as high energy density and long service life, an in-depth physical understanding of the complex battery processes is essential. The coupling of numerous complicated physico-chemical effects makes gaining insights into the behavior of batteries challenging for the scientific community. However, the increasing capabilities of machine learning (ML) algorithms point to a possible way to tackle this problem.

ML references a broad class of numerical or statistical algorithms for automatized data analysis, pattern recognition, classification, and regression. With increased computational power, ML techniques become more popular and researched. In battery research, there are many ML applications \cite{Lombardo2022}, e.g., early prediction of battery lifetime \cite{Severson2019}, material performance \cite{Jennings2019}, battery state estimation \cite{Hu2016, Li2021}, increase in simulation speed \cite{Botu2017}, and parameterization of physics-informed models \cite{Andersson2022, Bizeray2019, Jokar2016, Xu2022, Brady2020}. To make physical conclusions, analyzing available physical models and comparing them with experimental data utilizing ML methods is inevitable. Since this process is always accompanied by a lack of information, e.g., model uncertainty or measurement inaccuracies, a consistent uncertainty quantification (UQ) of the obtained results becomes increasingly important. The most natural incorporation of uncertainty is achieved by Bayesian algorithms, favoring them over other ML algorithms for these purposes. The capability of Bayes' theorem, updating your prior knowledge with more available evidence, and the results in multi-variate probability distributions make Bayesian algorithms a unique option in parameterization and inverse modeling. However, these algorithms usually need many simulated samples to obtain qualitatively good results \cite{Sethurajan2019, Aitio2020, Kim2023}, impeding the computationally heavy physical models. In this work, we show that using an improved Bayesian algorithm \emph{Expectation Propagation + Bayesian Optimization for Likelihood Free Inference} (EP-BOLFI) \cite{Kuhn2023} enables a good parameterization for physical models with UQ, needing orders of magnitude less samples in contrast to common Markov chain Monte Carlo (MCMC) methods. Further, the \emph{Bayesian alternately subsampled quadrature} (BASQ) model selection algorithm \cite{Adachi2023} based on Bayesian principles is applied to identify the prevailing mechanism from a specific selection. As a case study for these Bayesian algorithms, we investigate modeling solid-electrolyte interphase (SEI) growth, which limits the LiB lifetime.

Several degradation mechanisms \cite{Birkl2017, Edge2021} influence the lifetime of a LiB, e.g., the ongoing parasitic side reactions forming the solid-electrolyte interphase (SEI) \cite{Peled2017}. The SEI is crucial for the functionality of the lithium-ion battery by preventing direct contact between electrode and electrolyte \cite{Winter2009}. However, it does not ideally prevent side reactions as desired. The ongoing SEI growth consumes lithium ions and leads to capacity and performance losses \cite{Keil2016}. Despite long-lasting research efforts, the fundamental physical processes involved in the ongoing SEI growth are not fully understood \cite{Horstmann2019, Wang2018, Krauss2022}. Several mechanisms proposed in the literature aim to explain the continued SEI formation \cite{Single2018, Kolzenberg2020, Köbbing2023}. One can classify these mechanisms roughly into two categories: the first predicts that the prevailing mechanism is solvent diffusion through the SEI, and the second category postulates the SEI to be blocking for solvent molecules but considers ongoing electron transport through the SEI by various effects. These mechanisms can describe the capacity fade due to ongoing SEI formation but have subtle differences in the dependence on the operating conditions. These differences allow the disentangling of SEI growth mechanisms with ML approaches.

\begin{figure*}[]
	\centering
	\makebox[0pt]{%
    \includegraphics[width=0.85 \paperwidth]{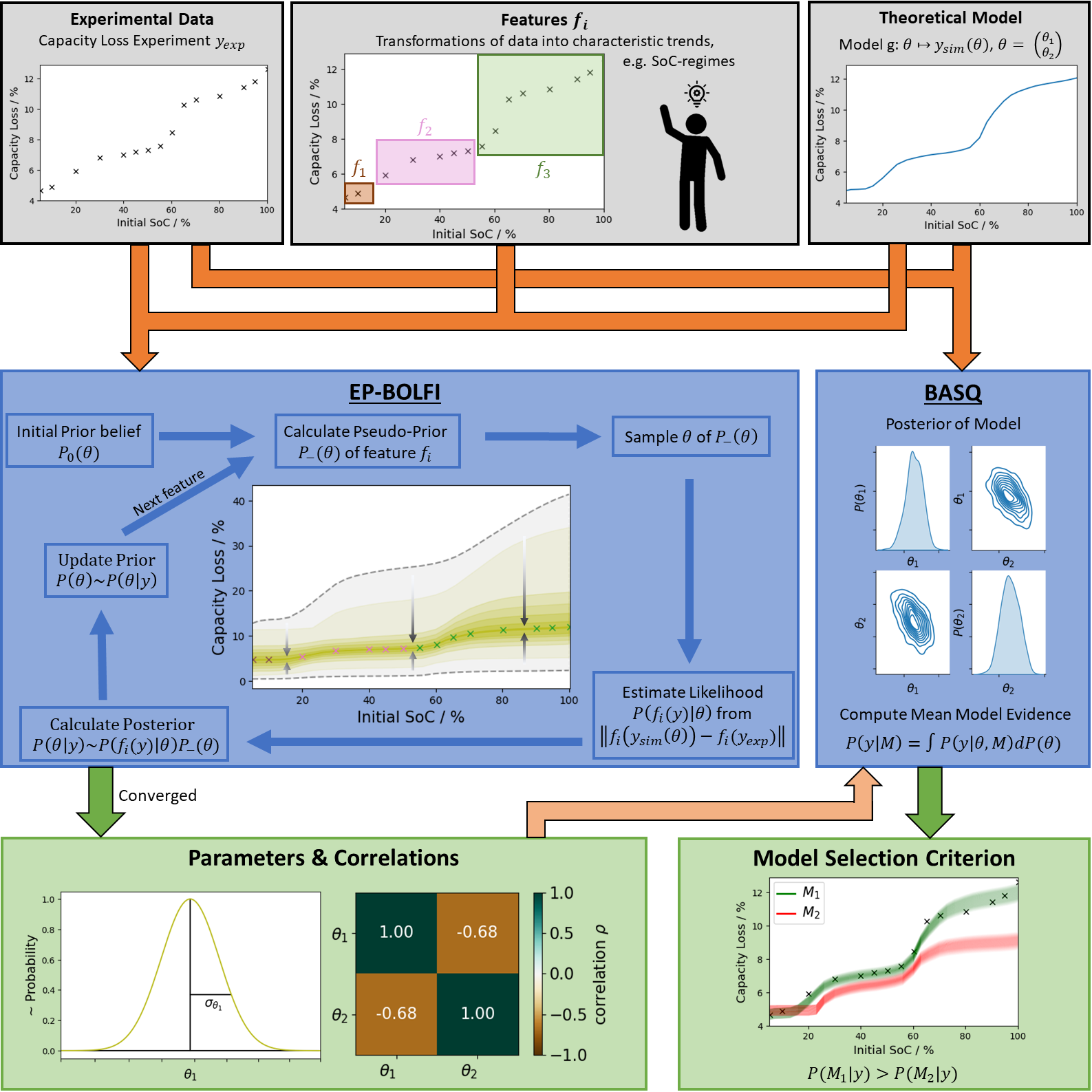}}
	\caption{Simplified visualization of the workflow presented in this work applied to the case study of modeling the continuous growth of the SEI. The upper row shows the primary input for the algorithms, i.e., the experimental data $y_\textrm{exp}$ (see Sec. \ref{subsec:ExpData}) and the model (e.g., a particular SEI growth mechanism, see Sec. \ref{subsec:SEIModels}). Human-chosen features are input to EP-BOLFI to improve its performance. The orange arrows indicate the input flows. The middle row shows the two Bayesian algorithms used in this work. EP-BOLFI performs a multidimensional probabilistic fit iteratively (see Sec. \ref{sec:EP-BOLFI}). BASQ (see Sec. \ref{sec:BQ}) determines the model posterior distribution and computes the mean model evidence. In the bottom row, the outputs are shown. From EP-BOLFI, the optimal parameter values and corresponding (co-)variances or correlations are obtained. To improve BASQ's performance, a preconditioned parameterization received by EP-BOLFI is helpful as additional input, as indicated by the faint orange arrow. Applying BASQ for different models yields a model selection criterion, as the "better" model obtains a higher mean model evidence.}
	\label{fig.WorkflowScheme}
\end{figure*}

This work presents a workflow for analyzing physically derived models with advanced Bayesian methods. We apply these methods to the example of battery aging due to continuous SEI formation. We present the workflow in Fig. \ref{fig.WorkflowScheme}. The Bayesian algorithms (see Sec. \ref{sec:BayesMethod}) take the experimental data and physics-based models (see Sec. \ref{sec:Methods}) as input and yield a parameterization with UQ, parameter correlations, and a model selection criterion as results (see Sec. \ref{sec:Results}). We conclude in Sec. \ref{sec:Conclusion}.

\section{Bayesian methods}\label{sec:BayesMethod}

In Bayesian statistics, the term probability is referred to certainty. In contrast to the usual frequentist approach, data points are considered fixed, and the parameters are uncertain. Instead of finding the perfect parameter value, the aim is to estimate the posterior certainty distribution of a desired parameter set $\theta$, considering the prior knowledge and the data $y$. Bayes' theorem states that given prior knowledge or assumption $P(\theta)$ can be updated with further knowledge or likelihood $P(y\mid \theta)$, to obtain a posterior distribution $P(\theta \mid y)$ by

\begin{equation}\label{eq.BayesTheroem}
    P(\theta \mid y) = \frac{P(y\mid \theta) P(\theta)}{P(y)},
\end{equation}
where $P(y)$ is the evidence of the data. In the following subsections, we introduce the EP-BOLFI and BASQ algorithms, which use Bayes' theorem and the Bayesian concept of probability.

\subsection{Bayesian Optimization}\label{sec:EP-BOLFI}

The algorithm used in this work for Bayesian Optimization is called Expectation Propagation + Bayesian Optimization for Likelihood Free Inference (EP-BOLFI), developed by \citet{Kuhn2023}. This algorithm is utilized for inverse modeling and combines two algorithms: Expectation Propagation (EP) and Bayesian Optimization for Likelihood Free Inference (BOLFI).

EP \cite{Minka2013} (see left blue box in Fig. \ref{fig.WorkflowScheme}) introduces the approach of splitting the data into several features $f_i$ and then propagating the gained information through every feature $f_{j \neq i}$, substantially reducing needed simulation samples. The choice of features is highly flexible and can be adapted to different problems. A trivial example is segmenting the data into different coherent parts and considering the data points in one single part as one feature. However, experts can also incorporate physical knowledge about the system by choosing certain data transformations as features, e.g., physically motivated fit functions. The feature then consists of the parameters describing the transformation. A feature can be thought of as a point in a higher-dimensional space. The resulting distance between this point for the simulated data and the point from the experimental data will be used for optimization. Therefore, it is essential to note that the selected transformations form the underlying landscape of a loss function in parameter space. Suitable choices can improve the results regarding noise stability, convergence speed, resulting uncertainty, and correct parameter identifiability.

BOLFI \cite{Gutmann2016} (see left blue box in Fig. \ref{fig.WorkflowScheme}) is responsible for fitting the selected features one by one to the experimental data. Firstly, different parameter configurations $\theta$ are drawn from the prior belief $P(\theta)$ and simulated. In the second step, the distance between the selected feature for the simulated data $f_i(y_\textrm{sim}(\theta))$ and the experimental data $f_i(y_\textrm{exp})$ is calculated. In the third step, a Gaussian process is trained on the parameter-distance pairs ($\theta, \textrm{log}(\vert \vert f_i(y_\mathrm{sim}(\theta)) - f_i(y_\mathrm{exp}) \vert \vert)$). Fourthly, transforming this Gaussian Process with 

\begin{equation}\label{eq:LL_threshold}
    P(y\mid \theta) \sim P(\textrm{log}(\vert \vert f_i(y_\mathrm{sim}(\theta)) - f_i(y_\mathrm{exp}) \vert \vert) \leq \epsilon),
\end{equation}
whereby $\epsilon$ is a certain threshold, yields the likelihood of the specific feature $f_i$. Fifthly, the product of the likelihood and the prior (see eq. (\ref{eq.BayesTheroem})) becomes evaluated by MCMC sampling. The result is the posterior distribution $P(\theta \mid y)$, without the need to explicitly determine the evidence $P(y)$.

Often, the likelihood is assumed to follow a particular distribution, e.g., a normal distribution. However, this may differ for coupled battery mechanisms, which can generate a more complex manifold in the parameter space. Therefore, the Gaussian process is used as a flexible surrogate for the likelihood.

The posterior is reduced to a multivariate normal distribution for better interpretability and ease in the following use. \citet{Barthelme2018} proved that in the case of a normal posterior, the EP procedure will converge to this exact posterior, and the reduction will not distort the result. A normal posterior is a good approximation, as identifiable problems tend to have a parabola-shaped optimum. In the sense of a traceback, the likelihood of the specific feature can then be calculated backwards from eq. (\ref{eq.BayesTheroem}), also as a normal distribution. The subsequent EP step realizes the propagation of the gained information. The obtained parameters' posterior distribution of feature $f_{i}$  enters as the prior belief for the next feature $f_{i+1}$. Besides, previously gathered information for this next feature ($f_{i+1}$) is removed from the prior. In this way, the prior belief contains only the information from the other already simulated features and/or the initial prior.

By iterating through the selected features (multiple times if needed), this procedure finally provides the overall posterior and feature-specific likelihoods as multivariate ellipsoids in hyperspace, with the most likely parameterization as the means and the corresponding uncertainty through the covariance matrices. Since the feature-specific likelihoods become updated during each iteration, we refer to the final likelihoods of the individual features as the feature-specific posteriors.

\subsection{Bayesian Quadrature}\label{sec:BQ}

Bayesian Quadrature (BQ) is a method to approximate intractable integrals. In the algorithm Bayesian alternately subsampled quadrature (BASQ, see right blue box in Fig. \ref{fig.WorkflowScheme}) developed by \citet{Adachi2023}, BQ is used to integrate the probability distribution $P(y \vert \theta, \textrm{M})$ of the data $y$ given the parameters $\theta$ and model $\textrm{M}$ over the parameter space of this model

\begin{equation}
    P(y \vert \textrm{M}) = \int P(y \vert \theta, \textrm{M})dP(\theta).
\end{equation}
With Bayes' theorem, one can relate this to a model probability 

\begin{equation}
    P(\textrm{M} \vert y) = \frac{P(y \vert \textrm{M}) P(\textrm{M})}{P(y)}.
\end{equation} 
Without further assumptions or information, all models enter with an equal prior probability $P(\textrm{M})$ and the same normalizing data evidence $P(y)$, leaving $P(y \vert \textrm{M})$ to be computed. Comparing the results for different models yields the Bayes factor
\begin{equation}
    K = \frac{P(\textrm{M}_1 \vert y)}{P(\textrm{M}_2 \vert y)} = \frac{P(y \vert \textrm{M}_1) P(\textrm{M}_1) P(y)}{P(y \vert \textrm{M}_2) P(\textrm{M}_2) P(y)} = \frac{P(y \vert \textrm{M}_1)}{P(y \vert \textrm{M}_2)}.
\end{equation}
Therefore, the model that achieves the highest value for the mean model evidence $P(y \vert \textrm{M})$ is considered the "best" model to describe the data. Note that this quantity itself is a non-normalized probability, only the comparison to a second model gives a significant meaning.

Considering the likelihood $P(y \vert \theta, \textrm{M})$ assigns a probability to a certain combination of model and parameter by weighting the distance to the data, this tends to prefer models with more available parameters to achieve the most accurate results. To avoid overfitting, we also investigate the effect of adding a penalty term to the likelihood, penalizing the number of parameters and their values (see SI \ref{sec:BASQ_Penalty}).


\section{Methods}\label{sec:Methods}

\subsection{Battery cell model}\label{subsec:CellModel}

Physics-informed battery models describe multiple coupled physico-chemical effects as partial differential equations (PDE). Even though resolving the physical effects in a battery most accurately requires a micro-structure-resolved 3D description \cite{Bolay2022}, numerically solving the system of PDEs in 3D is computationally heavy. Therefore, a trade-off between physical accuracy and computation time has to be made. The common reduction in complexity is reducing the system's dimensionality by volume-averaging. Thus, the Doyle-Fuller-Newman (DFN) \cite{Doyle1993} p2D model has been developed, which captures accurately most battery effects. However, it is still computationally challenging, especially for ML applications requiring thousands of simulations. Further reduction in complexity by asymptotic analysis of the DFN yields the Single Particle Model (SPM) and SPM considering electrolyte effects (SPMe) \cite{Marquis2019}. In this work, we will use the SPM/SPMe description of a battery.

In the simple SPM picture, all particles in each electrode act equally, so we only resolve one single particle. At the surface of this particle the Li-ions intercalate or deintercalate, and radially diffuse inside due to concentration gradients. This description is reasonable in the case of vanishing currents and, therefore, is used in this work to simulate battery storage. The SPMe model considers additional effects in the electrolyte, which is computationally slightly more expensive but more accurate for small applied currents. In this work, we use the SPMe to simulate the cycling behavior of the battery cell. The non-dimensional version of the governing equations of the SPMe, the model parameters, and the complementing initial and boundary conditions are summarized in SI \ref{sec:SPMe_Tables}. The battery simulations are performed with PyBaMM \cite{Sulzer2021}.

Commonly, the system is solved in time by initializing a certain battery state and giving the battery's current as input. The output contains all of the resolved battery states at different positions in space and time, in this work referred to as the simulated data $y_\textrm{sim}$. In this picture, we can describe the battery model as a function $g$, which takes the current $I$ and the battery parameters $\theta$ as input so that

\begin{equation}
    y_\mathrm{sim}(t) = g(I(t),\theta).
\end{equation}

For the case study pursued in this work (degradation by SEI formation), the most important simulated output is the battery's capacity loss (CL), which occurs due to SEI growth during battery operation/simulation and is described by the degradation models. For the parameterization, we assume to have an already correct parameterized underlying battery cell and focus on the parameterization of the degradation models only.

\subsection{Degradation models}\label{subsec:SEIModels}

In a battery, many degradation effects take place simultaneously. One of the least understood is the formation of the SEI, which is considered the dominant degradation process during battery storage but also contributes to degradation during battery operation. The circumstances in which multiple theoretical SEI models exist depict this as an ideal case study to use ML methods to identify which mechanisms take place and contribute up to which degree. Therefore, this is the focus of this work.

The SEI is a passivating layer at the negative electrode of a battery formed by the reduction reactions of electrolyte molecules. This reaction irreversibly consumes Li-ions, reducing the overall capacity and increasing the battery's internal resistance over time. To explain the observed growth of the SEI \cite{Huang2019} by ongoing reduction, a transport process of the reactants (electrons and solvent) to the reaction location has to take place. From a multi-scale perspective, the literature \cite{Single2018, Kolzenberg2020, Köbbing2023} proposes the following transport mechanisms for long-term SEI growth: electron diffusion, electron conduction or migration, and solvent diffusion. In the following, we introduce these mechanisms briefly.

Electron diffusion (ED) \cite{Single2017, Single2018, Soto2015, Shi2012, Shi2013, Kolzenberg2020} describes a diffusive transport of the electrons from the electrode through the SEI by hopping between localized states, e.g., lithium interstitials. At the interface between SEI and electrolyte these electrons are consumed in SEI formation reactions. How fast this reaction takes place depends on local quantities like the onset potential of the SEI formation reaction $\Phi_0$ and the electrical potential. \citet{Kolzenberg2020} investigated a reaction limitation in more detail and showed that this is important only for the early stages of SEI formation. Hence, we neglect the influence of the onset potential here and assume instantaneous reactions once all reactants arrive at the reaction site, i.e., SEI formation reactions instantly consume all electrons. This leads to a vanishing concentration of electrons at the interface between SEI and electrolyte and a specific reference concentration at the electrode, which depends on the state of charge (SoC). Then, the diffusive transport of electrons that contribute to the SEI reaction is given by the following electron current density

\begin{equation}\label{eq.j_ED}
    j_\mathrm{ED} = \frac{c_\mathrm{e^-}D_\mathrm{e^-}F}{L_\mathrm{SEI}}\mathrm{exp}\left(-\tilde{\eta}_\mathrm{SEI}\right),
\end{equation}
where $c_\mathrm{e^-}$ is the reference concentration, $D_\mathrm{e^-}$ is the diffusion constant of electrons through such localized states, $F$ is the Faraday constant, and $L_\mathrm{SEI}$ is the thickness of the SEI. $\tilde{\eta}_\mathrm{SEI}$ is the SEI overpotential, which describes the amount of electrons available in the SEI at the electrode-SEI interface. This is given as

\begin{equation}
    \tilde{\eta}_\mathrm{SEI} = \frac{F}{RT}\left(\eta_\mathrm{int} + U(\mathrm{SoC}) + \frac{\mu_\mathrm{Li,0}}{F}\right),
\end{equation}
where $R$ is the universal gas constant, $T$ is the temperature, $\mu_\mathrm{Li,0}$ is the standard chemical potential of neutral lithium to lithium metal, $U(\mathrm{SoC})$ is the open-circuit voltage of the anode, and $\eta_\mathrm{int}$ is the intercalation overpotential. For standard Butler-Volmer kinetics, this intercalation overpotential is defined by the following relation to the intercalation current density

\begin{equation}
    j_\mathrm{int} = 2 j_\mathrm{0} \mathrm{sinh}\left(\frac{F}{2RT}\eta_\mathrm{int}\right),
\end{equation}
with the exchange current density $j_\mathrm{0}$.

\begin{figure}[]
	\centering
	\makebox[0pt]{%
    \includegraphics[width=0.43\paperwidth]{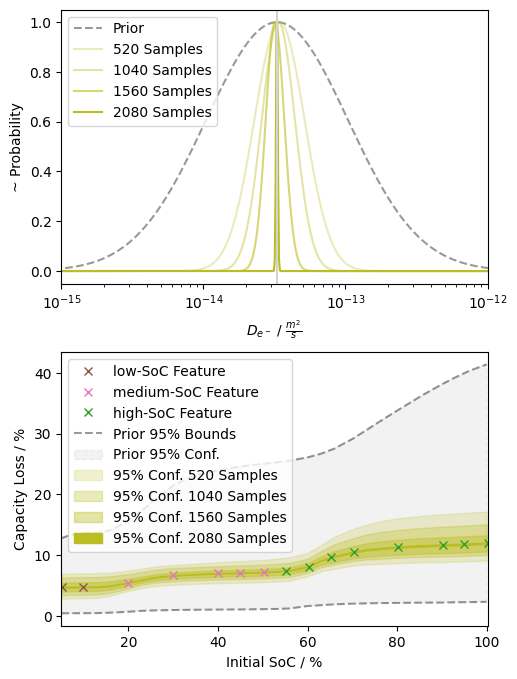}}
	\caption{Exemplary evolution of one parameter's probability distribution (top panel) with simulated samples. The dashed line indicates the prior belief for this parameter. The colored lines represent the sequence of probability distributions after 520 simulated samples (transparent) up to 2080 samples (opaque). For better comparison, the probability distributions are normalized to one. The vertical grey line indicates the real parameter value for the synthetic data. In the lower panel, the corresponding 95\% confidence areas of the parameterization are shown in data space. The crosses refer to the synthetic data points, and their coloring indicates their featurization into three segments: low SoC, medium SoC, and high SoC.}
	\label{fig.SynthStor_Par+UQ_Evo}
\end{figure}

\begin{figure*}[]
	\centering
	\makebox[0pt]{%
    \includegraphics[width=0.85\paperwidth]{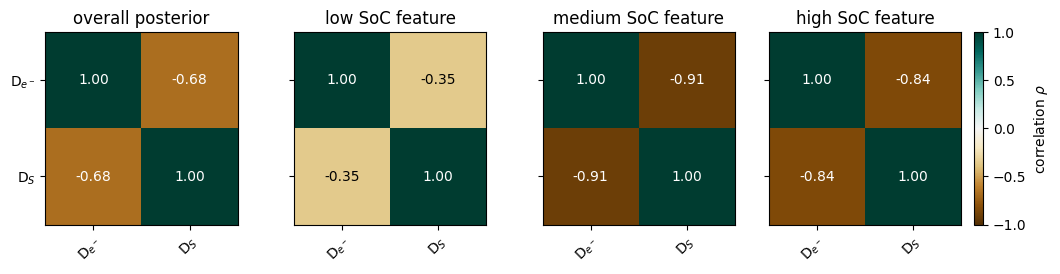}}
	\caption{Correlation coefficients between the parameters $D_\mathrm{e^-}$ and $D_\mathrm{S}$ for the inverse modeling of the synthetic storage data with the "Best Model". The leftmost panel shows the correlation in the overall posterior. The other panels refer to the feature-specific correlations in the low, medium, and high SoC features. The color indicates the value of the correlation coefficients.}
	\label{fig.SynthStor_Correlations}
\end{figure*}

Another possible transport mechanism of electrons can occur due to a gradient in the electrical potential through the SEI. Combined with Ohms's law, this leads to a net current of electrons from the anode to the SEI-electrolyte surface. Depending on the origin of the gradient in the electrical potential, this can yield different current densities. One possibility, referred to as electron conduction (EC) \cite{Single2018, Horstmann2019}, assumes a linear gradient in the electrical potential over the thickness of the SEI during storage. Its current density is given by

\begin{equation}\label{eq.j_EC}
    j_\mathrm{EC} = \frac{\kappa (\Phi_0 - U(\mathrm{SoC}))}{L_\mathrm{SEI}},
\end{equation}
where $\kappa$ is the electron conductivity of the SEI, and $\Phi_0$ is the onset potential of the SEI formation reaction. The solvent becomes unstable at this potential and becomes reduced to form SEI compounds. Note that this only provides a current for SEI formation if $U(\mathrm{SoC}) < \Phi_0$. The leading critics of this mechanism capture the conductivity of the SEI, which is considered to be an insulator in the relevant voltage regime \cite{Xu2023, Chen2011}, and the assumption of a linear gradient in the electrical potential from the anode to the SEI-electrolyte interface. The electrical potential drops at the interfaces between electrode and SEI and between SEI and electrolyte, putting the assumption of one single linear gradient over the whole range into question \cite{Lück2018}.

During battery operation, especially charging, the intercalation current of positively charged Li-ions causes a change in the electrical potential at the electrode-SEI interface \cite{Kolzenberg2020}. The assumption that this electrical potential drops towards the SEI-electrolyte interface following a linear gradient, referred to as electron migration (EM), causes the following current density

\begin{equation}\label{eq.j_EM}
    j_\mathrm{EM} = \frac{c_\mathrm{e^-}D_\mathrm{e^-}F^2j_\mathrm{int}}{2RT\kappa_\mathrm{{Li}^+,SEI}}\mathrm{exp}\left(-\tilde{\eta}_\mathrm{SEI}\right),
\end{equation}
where $\kappa_\mathrm{{Li}^+,SEI}$ is the lithium-ion conductivity of the SEI.

Solvent diffusion (SD) \cite{Ploehn2004, Ramasamy2007, Sankarasubramanian2012, Pinson2013, Hao2017} assumes that the continuous transport of solvent limits the reduction reactions. One assumes a vanishing concentration of the solvent at the reaction site, again due to instantaneous reactions, and a constant solvent concentration in the bulk electrolyte. Then, the transport of the solvent is realized by diffusion, due to concentration gradients, through the SEI to the electrode reaction surface. Deriving a current density, in means of electrons lost to SEI formation, this mechanism yields

\begin{equation}\label{eq.j_SD}
    j_\mathrm{SD} = \frac{c_\mathrm{S}D_\mathrm{S}F}{L_\mathrm{SEI}},
\end{equation}
where $c_\mathrm{S}$ is the concentration of solvent molecules in bulk, and $D_\mathrm{S}$ is the diffusion constant of the solvent molecules through the SEI. The physical credibility behind the process is also highly discussed, e.g., the solvent molecules have to be able to diffuse through the SEI pores, even though their reaction should close those \cite{Krauss2024}.

Despite decades-long research efforts, the scientific community cannot fully explain the SEI growth, suggesting that the SEI is complex and possibly based on multiple coupled mechanisms. \citet{Single2018} and \citet{Köbbing2023} investigated SEI growth under storage conditions and found that electron diffusion best describes the SoC-dependent trend in the experimental data. However, both works need an additional SoC-independent capacity loss, whose origin is unclear. The SOC-independent capacity loss is possibly due to an additional SEI growth mechanism, where only solvent diffusion matches this functionality. Therefore, we model SEI growth by combining SoC-dependent mechanisms (electron diffusion or electron conduction) with solvent diffusion. With this knowledge, we investigate the following combinations as SEI models for battery storage. Firstly, we label the combination of electron diffusion and solvent diffusion as the "Best Model" since previous work verified that it matches the experimental data. Secondly, we label the combination of electron conduction (with $\Phi_0 =0.145$V) and solvent diffusion as the "Wrong model" because the parameter value needed for $\Phi_0$ to capture the trends in the experimental data is unreasonable. Thirdly, we consider the combination of electron diffusion, electron conduction, and solvent diffusion, which we label the "Overfitted Model" because it has more fit parameters than needed. For battery cycling, we additionally have to include electron migration. We label the combination of electron diffusion, solvent diffusion, and electron migration as the "Cycling Model".

To model such combinations, we combine the derived current densities linearly to a total SEI current density $j_\mathrm{SEI} = \sum j_\mathrm{i}$, in case multiple transport mechanisms occur at a time. This current density is then incorporated into the battery model by assuming that the overall battery cell current density $j$ consists of the intercalation current density $j_\mathrm{int}$ and the SEI current density $j_\mathrm{SEI}$ so that $j = j_\mathrm{int} + j_\mathrm{SEI}$. 

Depending on the mean molar volume of the formed SEI molecules, $V_\mathrm{SEI}$, the SEI thickness grows in time by

\begin{equation}\label{eq:dL_SEIdt}
    \frac{\mathrm{d}L_\mathrm{SEI}}{\mathrm{d}t} = \frac{V_\mathrm{SEI}}{F}j_\mathrm{SEI}.
\end{equation}
The solutions of this equation for single transport processes are shown in SI \ref{sec:SEI_thickness}. Note that these solutions strongly differ in their SoC dependency \cite{Single2018}.

The proposed SEI growth mechanisms depend on many parameters, often appearing as direct products, making the parameters for a fitting purpose indistinguishable. Therefore, we choose many parameters to be fixed and usually perform the parameterization with only one parameter per mechanism, granting a faster convergence. Then, each mechanism is characterized by a single parameter. As a consequence, the correlations between these parameters can also be interpreted as the relations between the degradation mechanisms. Then, the remaining parameters of interest are $D_\mathrm{e^-}$, $\kappa$, $D_\mathrm{S}$, and $\kappa_\mathrm{{Li}^+,SEI}$.

\subsection{Experimental Data}\label{subsec:ExpData}

As outlined in the previous section, one characteristic of the theoretical SEI degradation mechanisms is their dependence on the SoC. To investigate this dependence and possibly identify the actual SEI transport mechanisms, we inversely analyze the experimental data obtained by \citet{Keil2016}. Here, a very brief outline of the experiment is given. In the experiment, they investigated the capacity loss of lithium-ion batteries with a nickel-cobalt-aluminium oxide (NCA) cathode. They charged batteries to different initial SoCs and stored them for 9.5 months, performing check-ups about every two months. As a result, they measured a clear trend of the capacity loss with time and SoC: the higher the initial SoC, the higher the capacity loss increasing in time. Notably, they detected a significant degradation (about 4 \%) for 0 \% initial SoC. By measuring the open-circuit voltage $U$ of the graphite anode, it is possible to model the measured capacity loss inversely with the proposed degradation mechanisms.

\subsection{Synthetic Data}\label{subsec:SynthData}

To check whether EP-BOLFI and BASQ work as intended, we produce synthetic data sets for which we know the occurring degradation mechanisms and their correct parameterization.

As a first synthetic dataset, we simulate SEI growth with the "Best Model" during the experimental storage protocol described in Sec. \ref{subsec:ExpData}. Therefore, the synthetic storage data (see crosses in the lower panel of Fig. \ref{fig.SynthStor_Par+UQ_Evo}) is uniquely parameterized by the correct values for $D_\mathrm{e^-}$ and $D_\mathrm{S}$.

Additionally, we produce a second synthetic dataset to show the application of the methods on cycling data, which is a widespread experimental procedure. Further, we use this dataset to investigate EP-BOLFI's performance under more difficult conditions and analyze the impact of suitable feature choices on performance and obtained results. We generate the synthetic cycling data (see green line in Fig. \ref{fig.Cycl_ConvEvo}) by simulating SEI growth with the "Cycling Model" for 500 full cycles with 1C CC-CV charge and 1C constant discharge. The corresponding voltage cut-offs are 2.5V and 4.2V, with a cut-off current of C/50. If noise does not exist, this data is uniquely parameterized with correct values for $D_\mathrm{e^-}$, $D_\mathrm{S}$ and $\kappa_\mathrm{{Li}^+,SEI}$. We then apply different noise levels to this data for the analysis.

\section{Results \& Discussion}\label{sec:Results}

\begin{figure*}[]
	\centering
	\makebox[0pt]{%
    \includegraphics[width=0.85\paperwidth]{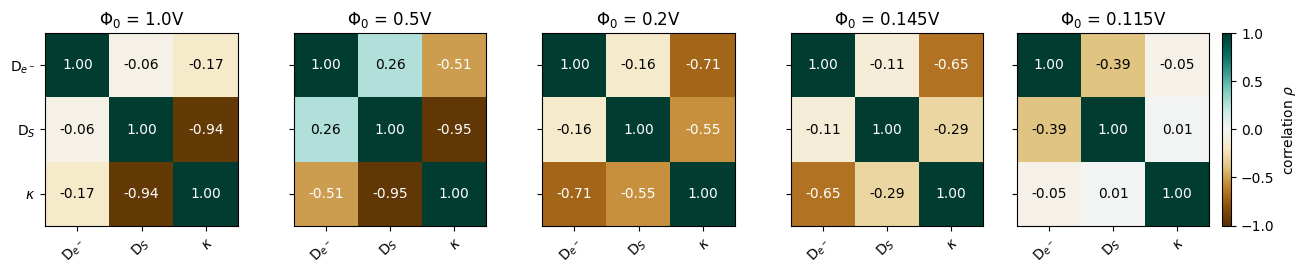}}
	\caption{Correlation coefficients $\rho$ between the parameters $D_\mathrm{e^-}$, $D_\mathrm{S}$, and $\kappa$ for the inverse modeling of the synthetic storage data with the "Overfitted Model". The different panels correspond to five different values of the onset potential $\Phi_0$. From left to right, $\Phi_0$ decreases, mimicking a transition of the electron conduction mechanism from SoC-independent ($\Phi_0 \in \{1.0 \textrm{V}, 0.5 \textrm{V} \}$) to SoC-dependent ($\Phi_0 \in \{0.2 \textrm{V}, 0.145 \textrm{V} \}$) to a vanishing contribution ($\Phi_0 = 0.115 \textrm{V}$). The color indicates the value of the correlation coefficients.}
	\label{fig.SynthStor_CorrTrans}
\end{figure*}

In this section, we show the results of applying the presented Bayesian methods to model the degradation data inversely with SEI growth. In the first subsection, we analyze storage data with EP-BOLFI and BASQ to verify the methods and identify the dominant transport mechanism. In the second subsection, we show the results of using EP-BOLFI to analyze cycling data, a ubiquitous experimental protocol. We investigate the impact of noise and the choice of features on the results. We discuss our findings along the way.

\subsection{Storage Data}

We first present the achieved results for synthetic storage data (see Sec. \ref{subsec:SynthData}) to demonstrate the capability and credibility of the Bayesian methods. Afterwards, the results for real storage data (see Sec. \ref{subsec:ExpData}) are shown and discussed.

For the inverse analysis of the storage data with EP-BOLFI, features and a prior belief of the parameter range are needed. To enable correct identification of $D_\mathrm{S}$ (see eq. \ref{eq.j_SD}), which is responsible for the low SoC contribution, and $D_\mathrm{e^-}$ (see eq. \ref{eq.j_ED}) or $\kappa$ (see eq. \ref{eq.j_EC}), which are responsible for the SoC-dependent capacity loss, we choose the features to split the data into three SoC segments (see colored crosses in the lower panel). The 95\% bounds of the prior belief for the parameters are chosen two orders of magnitudes around the correct/guessed value, such that the lower (higher) limit for each of the corresponding transport mechanisms would cause significantly less (more) degradation than observed in the data.

The results of the parameterization of the synthetic storage data with the "Best Model" are shown in Fig. \ref{fig.SynthStor_Par+UQ_Evo}. The upper panel of Fig. \ref{fig.SynthStor_Par+UQ_Evo} shows the prior belief and sequential convergence of the probability distribution to the correct value with increasing knowledge (simulated samples), exemplary for $D_\mathrm{e^-}$. In the lower panel of Fig. \ref{fig.SynthStor_Par+UQ_Evo} the simulations with the parameter values at the bounds of the sequential 95\% areas for $D_\mathrm{e^-}$ and $D_\mathrm{S}$ are shown. Here, the colored areas refer to the 95\% confidence areas of the joint probability distribution, again at different states of the algorithm. The more samples are analyzed, the narrower the confidence areas become around the synthetic data.
Without considering noise, the data is uniquely parameterized by known values for $D_\mathrm{e^-}$ and $D_\mathrm{S}$. EP-BOLFI finds the correct parameterization to describe the synthetic data perfectly. The uncertainty vanishes, and the solution converges within multiple thousand samples, even for a wide prior parameter space. Note that the needed samples can be drastically reduced by ideally adapting EP-BOLFI's hyperparameters (e.g., weaker dampening).

Besides the final parameterization, i.e., the means and variances of the final joint probability distribution, EP-BOLFI also outputs the covariances. From these, the Pearson correlation coefficients $\rho(X,Y)$ for the optimized parameters $X$ and $Y$ can be calculated. The correlation coefficient measures the linear connection between $X$ and $Y$. For parameterization, the correlation between two parameters can be understood as to to what degree parameter $X$ has to be modified if $Y$ changes to get a similar quality to fit the trend in the data. In reverse, the correlation coefficient contains the information on whether a specific parameter can be identified independently, i.e., has vanishing correlations to other parameters. As we describe each degradation mechanism with one parameter only, the correlation coefficient measures how identical or exchangeable these mechanisms are.
Figure \ref{fig.SynthStor_Correlations} shows the correlation coefficients between the parameters $D_\mathrm{e^-}$ and $D_\mathrm{S}$ for the overall posterior and the specific features obtained in the inverse modeling of the synthetic data. In the low SoC feature, the parameters show a low anti-correlation, as solvent diffusion dominates the SEI growth. A strong anti-correlation is achieved in the medium SoC feature, due to an almost equal contribution of electron diffusion and solvent diffusion. In the high SoC feature, the anti-correlation is slightly reduced as electron diffusion dominates the SEI growth in this regime. The overall posterior yields an anti-correlation of $\rho = -0.68$, which is, in fact, a combination of the prior correlation and the correlations in the selected features. Therefore, the value lies between the others, as expected. For a more detailed analysis and discussion of feature-specific correlations, see SI \ref{sec:Feat_Corr}.

\begin{figure*}[]
	\centering
	\makebox[0pt]{%
    \includegraphics[width=0.85\paperwidth]{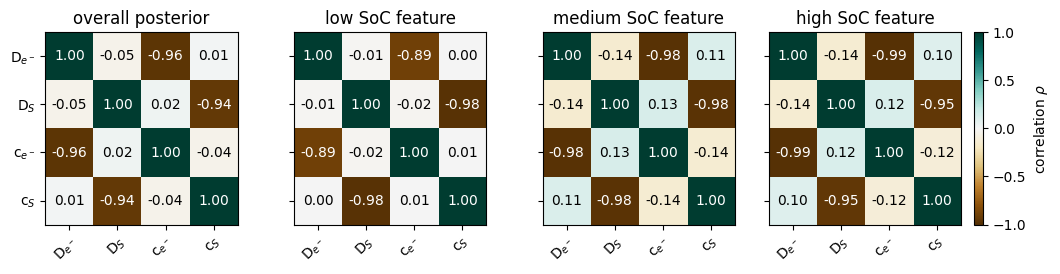}}
	\caption{Correlation coefficients $\rho$ between the parameters $D_\mathrm{e^-}$, $D_\mathrm{S}$, $c_\mathrm{e^-}$, and $c_\mathrm{S}$ for the inverse modeling of the synthetic storage data with the "Best Model". The leftmost panel shows the correlations in the overall posterior. The other panels refer to the feature-specific correlations in the low, medium, and high SoC features. The color indicates the value of the correlation coefficients.}
	\label{fig.SynthStor_Corr4P}
\end{figure*}

To further investigate whether the obtained correlation values of the overall posterior reflect the physically expected interplay between the degradation mechanisms, the synthetic data is inversely modeled with the "Overfitted Model". By varying the value for $\Phi_0$, the electron conduction mechanism changes its SoC dependence. Note that experimental measures reveal that values for $\Phi_0$ lie between 0.8V and 1.5V \cite{Xu2004}. However, we also use lower values to mimic a similar SoC dependency as electron diffusion. Figure \ref{fig.SynthStor_CorrTrans} shows the overall posterior correlations between $D_\mathrm{e^-}$, $D_\mathrm{S}$, and $\kappa$ for five different $\Phi_0$-values, decreasing from left to right. $\rho(D_\mathrm{S},\kappa)$ shows a direct anti-correlation for high values of $\Phi_0$. It decreases for $\Phi_0 < 0.5$V and vanishes for $\Phi_0 = 0.115$V. Simultaneously, the anti-correlation of $D_\mathrm{e^-}$ and $\kappa$ increases to $\rho(D_\mathrm{e^-},\kappa) = -0.71$ as the onset potential decreases to $\Phi_0 = 0.2$V. For lower values of $\Phi_0$, $\rho(D_\mathrm{e^-},\kappa)$ decreases or vanishes.
The obtained correlation values behave as expected. For $\Phi_0 \in \{1.0 \textrm{V}, 0.5 \textrm{V} \}$, electron conduction leads to almost constant capacity loss over the SoC range, similar to solvent diffusion. Therefore, a strong anti-correlation is obtained for $\rho(D_\mathrm{S},\kappa) \approx -1$, meaning these processes behave here very similarly, and only a weak anti-correlation for $\rho(D_\mathrm{e^-},\kappa)$. This changes for $\Phi_0 \in \{0.2 \textrm{V}, 0.145 \textrm{V} \}$ because then electron conduction contributes stronger and only for higher SoC-values, becoming similar to electron diffusion. For $\Phi_0 = 0.115$V, electron conduction only contributes to the highest SoC points, resulting in vanishing overall correlation with the other mechanisms.

Despite the interplay and exchangeability between the physical processes, the correlation values can also indicate an overparameterized model. To check if EP-BOLFI can find lumped parameters, we inversely model the synthetic data with the "Best Model". Here, we optimize the parameters $D_\mathrm{e^-}$, $D_\mathrm{S}$, $c_\mathrm{e^-}$, and $c_\mathrm{S}$, where only the product of $D_\mathrm{e^-}$ and $c_\mathrm{e^-}$ (see eq. \ref{eq.j_ED}) and $D_\mathrm{S}$ and $c_\mathrm{S}$ (see eq. \ref{eq.j_SD}) are important. The resulting correlations are shown in Fig. \ref{fig.SynthStor_Corr4P}. EP-BOLFI finds a clear anti-correlation between these parameter pairs ($\rho(D_\mathrm{e^-},D_\mathrm{e^-})$ and $\rho(D_\mathrm{e^-},D_\mathrm{e^-})$) in all features, as expected since the parameters appear as direct products.

\begin{figure}[]
	\centering
	\makebox[0pt]{%
    \includegraphics[width=0.43\paperwidth]{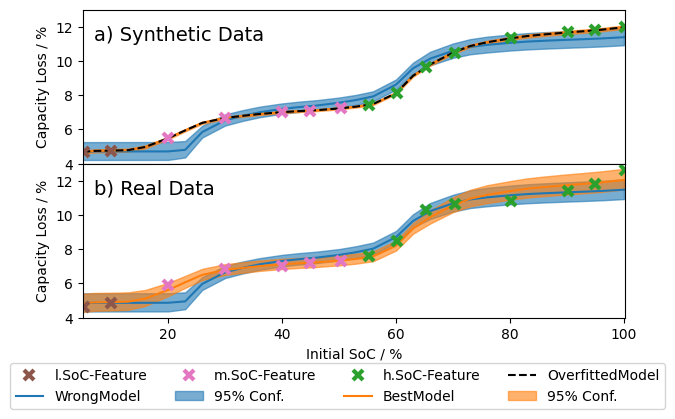}}
	\caption{Inverse modeling of a) synthetic storage data and b) real storage data with the "Best Model" (orange) and "Wrong Model" (blue). The colored areas show the 95\% confidence area of the parameterization. The colored crosses indicate the featurization of the data. The parameterization of the "Overfitted Model" with $\Phi_0 =0.145$V (dashed black) is also shown for the synthetic storage data.}
	\label{fig.SynthStor_Model_Comparison}
\end{figure}

\begin{table*}[]
\centering
\begin{tabular}{||c || c | c | c | c||} 
 \hline
 Models & Synt. Data & Synt. Data + Penalty & Real Data & Real Data + Penalty\\ [0.5ex] 
 \hline\hline 
 "Best Model" & 183.68 & \textbf{178.67} & 2.14 & \textbf{-9.12} \\ 
 "Wrong Model" & -130.19 & -144.35 & -282.19 & -291.42 \\
 "Overfitted Model" ($\Phi_0 =0.145$V) & \textbf{187.24} & 168.34 & \textbf{3.95} & -13.13 \\ [0.5ex] 
 \hline
\end{tabular}
\caption{Mean model evidence calculated by BASQ for synthetic data and real storage data with and without penalty (due to higher number of fit parameters). The number in bold indicate the model with the highest mean model evidence.}
\label{tab.BASQ_Storage}
\end{table*}

Figure \ref{fig.SynthStor_Model_Comparison} shows the results of inversely modeling the synthetic (upper panel) and real storage data (lower panel) with different models. The parameterized models "Best Model" (orange) and "Wrong Model" (blue) are shown with their corresponding 95\% confidence intervals after equal sample numbers. In the upper panel, the final parameterization of the "Overfitted Model" (with $\Phi_0 = 0.145$V) is visualized by the dashed black line and coincides with the curve of the "Best Model". Even though the correct model perfectly describes the synthetic data, the "Wrong Model" catches a similar trend in the SoC behavior. Without a mathematically reliable framework, it is difficult to evaluate which of the models fits the data the best.
The parameterized models for the real storage data are shown in the lower panel. Again, both combinations follow the trend in the data reasonably well, and the corresponding 95\% confidence areas include almost every data point. This demands a suitable model selection criterion that also considers the confidence areas.

Therefore, we use the BASQ algorithm to investigate the models, and its results are shown in Tab. \ref{tab.BASQ_Storage}. The bold numbers indicate which model achieved the highest mean model evidence computed by BASQ in the different analyses. BASQ identifies the "Best Model" as better than the "Wrong Model" for the synthetic and real data. However, the "Overfitted Model" achieves slightly higher mean model evidence. As seen before, the best parameterization for the "Overfitted Model" and "Good Model" coincides visually (upper panel of Fig. \ref{fig.SynthStor_Model_Comparison}), so the higher evidence is assumed to be due to overfitting. Two options emerge to cope with the challenge of overparameterization: firstly, consider only reasonable parameter ranges. This would directly lead to the deselection of the models "Wrong Model" and "Overfitted Model" due to the nonphysical values of $\Phi_0$. The second option is introducing a penalty, which we pursue in this work (see SI \ref{sec:BASQ_Penalty}). This penalty lowers the likelihood of a certain parameter combination depending on the amount of parameters used in the model. Thus, a model with more optional parameters becomes less likely (i.e., achieves lower evidence) if it does not better fit the data. Then, the mean model evidence results, including a penalty, identify "Best Model" as the best model for synthetic data, as expected, and as the best model for the real storage data. Since solvent diffusion was used in this work to represent the SoC-independent degradation and explain the capacity loss at low SoCs, BASQ identifies electron diffusion as the best transport mechanism to describe the SoC-dependent characteristics in this storage data.

\begin{figure}[]
	\centering
	\makebox[0pt]{%
    \includegraphics[width=0.43\paperwidth]{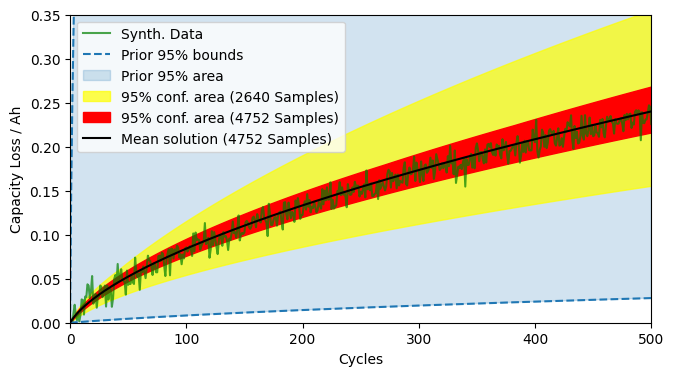}}
	\caption{Inverse modeling of the synthetic data ($\sigma_\textrm{noise}^2 = 8 \cdot 10^{-5} \textrm{Ah}^2$) with the correct "Cycling Model". With an increasing number of drawn samples, the parameters' joint probability distribution converges to the correct parameterization (black line) and a reasonable 95\% confidence area (red) considering the noise level. The initial prior is wide (indicated by the blue 95\% prior area) and significantly biased (non-symmetric around the true values for the parameters).}
	\label{fig.Cycl_ConvEvo}
\end{figure}

\subsection{Cycling Data}

In this subsection, the results for analyzing synthetic cycling data (see Sec. \ref{subsec:SynthData}), as a prevalent experimental protocol, with EP-BOLFI under impeded conditions (significant bias, wide prior belief, and noisy data) are shown. The obtained uncertainty, its dependence on noise ratio, and chosen featurization are investigated and discussed.

\begin{table}[]
\centering
\begin{tabular}{||c || c | c | c||} 
 \hline
 Parameter & True Value & Mean Prior Value & Final Fit Value\\ [0.5ex] 
 \hline\hline 
 $D_\mathrm{e^-}$ & $3.3 \cdot 10^{-14} \frac{\textrm{m}^2}{\textrm{s}}$ & $7.0 \cdot 10^{-14} \frac{\textrm{m}^2}{\textrm{s}}$ & $3.36 \cdot 10^{-14} \frac{\textrm{m}^2}{\textrm{s}}$  \\ 
 $D_\mathrm{S}$ & $2.5 \cdot 10^{-21} \frac{\textrm{m}^2}{\textrm{s}}$ & $8.0 \cdot 10^{-21} \frac{\textrm{m}^2}{\textrm{s}}$ & $2.53 \cdot 10^{-21} \frac{\textrm{m}^2}{\textrm{s}}$  \\
 $\kappa_\mathrm{{Li}^+,SEI}$ & $1.0 \cdot 10^{-6} \frac{\textrm{S}}{\textrm{m}}$ & $8.0 \cdot 10^{-6} \frac{\textrm{S}}{\textrm{m}}$  & $0.99 \cdot 10^{-6} \frac{\textrm{S}}{\textrm{m}}$ \\ [0.5ex] 
 \hline
\end{tabular}
\caption{Comparison of parameter values for the inverse modeling of the synthetic cycling data. The true values correspond to the values used to produce the synthetic data. The mean prior values indicate the biased prior belief by placing the mean of the prior parameter distributions to bad "guessed" values. The final fit values are the means of the overall posterior distribution obtained after 4752 samples and coincide with the mean solution plotted in Fig. \ref{fig.Cycl_ConvEvo}.}
\label{tab.Cycl_ParValues}
\end{table}

Again, EP-BOLFI needs features and a prior belief for the parameters. For featurization, we have chosen a total of two different features. In the first feature, the data is fitted by a power law function ($y(t) \sim \alpha t^{\beta}$) to capture the correct contributions of the degradation mechanisms and thereby grant the proper trajectory of the capacity loss. Then, the first feature is $f_1 = [\alpha, \beta]^{\mathrm{T}}$. In the second feature, the capacity loss averaged over the last ten cycles is chosen $f_2 = [\overline{\textrm{CL}}(t_\mathrm{av})]$ to capture an absolute value of the capacity loss at the end of the cycling. The parameters' prior distributions are biased by setting the means to values two to eight times larger than the correct values used for the synthetic data (see Tab. \ref{tab.Cycl_ParValues}). The width of the prior distributions is chosen such that the 95\% confidence bounds are one order of magnitude lower or higher, respectively, than the mean value.  

Figure \ref{fig.Cycl_ConvEvo} shows the resulting sequential convergence in the parameterization of the synthetic cycling data (green line) with the correct "Cycling Model" for a relatively high noise level ($\sigma_\textrm{noise}^2 = 8 \cdot 10^{-5} \textrm{Ah}^2$). The light blue area indicates the 95\% confidence area of the initial prior belief. With increasing knowledge (simulated samples), the 95\% confidence bounds of the joint probability hyper-ellipsoid in data space (indicated by the colored areas) become symmetric around the data, and the uncertainty converges to a reasonable level. More simulated samples do not decrease the uncertainty significantly as it represents the noise in the data. E.g., the 95\% confidence area after 6800 samples is very similar to the one after 4752 samples. Table \ref{tab.Cycl_ParValues} shows the parameter values for the simulated synthetic data, the initial prior means, and the final parameters. Note that these parameter values depend on other battery parameters and, therefore, should not be taken as absolute, i.e., a falsely assumed negative electrode surface area can change these parameters by orders of magnitude. The focus here is to show the correct identification of the "true" synthetic parameters. EP-BOLFI detects a lower impact of solvent diffusion as initially guessed and, therefore, reduces $D_\mathrm{S}$ to a value that shows an almost vanishing contribution of solvent diffusion. Equivalently, the value for $D_\mathrm{e^-}$ is lowered to the correct value, which is a more difficult task since $D_\mathrm{e^-}$ contributes to both $j_\mathrm{ED}$ (eq. (\ref{eq.j_ED})) and $j_\mathrm{EM}$ (eq. (\ref{eq.j_EM})). This means $D_\mathrm{e^-}$ describes the correct contribution of a square root-like capacity loss and influences additionally the linear capacity loss over time. A change in $D_\mathrm{e^-}$ will affect both. However, $\kappa_\mathrm{{Li}^+,SEI}$ affects the linear aspect only (see eq. (\ref{eq.j_EM})). Identifying the right combination of $D_\mathrm{e^-}$ and $\kappa_\mathrm{{Li}^+,SEI}$ represents the correct interplay of a square root and linear SEI growth regime. The relative error of the parameters is below 2\%, even for this noise level. One can see that even for a wrong initial guess, a very wide prior (blue area in Fig. \ref{fig.Cycl_ConvEvo}), and a significant noise level, a valid parameterization with reasonable uncertainty is found by EP-BOLFI with a low number of samples.

Table \ref{tab.Cov_NoiseLevel} shows the posterior variances for the parameters of the "Cycling Model", obtained after inversely modeling the synthetic cycling data with different applied noise levels. The lower the noise level, the lower the obtained parameter variances. The uncertainty quantification of EP-BOLFI indeed captures the noise and uncertainty in the data by the resulting confidence in the parameter space. 

\begin{table}[]
\centering
\begin{tabular}{||c || c | c||} 
 \hline
 Variances & $\sigma_\textrm{noise}^2 = 8 \cdot 10^{-6} \textrm{Ah}^2$ & $\sigma_\textrm{noise}^2 = 8 \cdot 10^{-5} \textrm{Ah}^2$\\ [0.5ex] 
 \hline\hline 
 $\sigma_{D_\mathrm{e^-}}^2$  & 0.01 & 0.08 \\ 
 $\sigma_{D_\mathrm{S}}^2$  & 0.24 & 0.45 \\
 $\sigma_{\kappa_\mathrm{{Li}^+,SEI}}^2$  & 0.02 & 0.13 \\ [0.5ex] 
 \hline
\end{tabular}
\caption{Non-dimensional posterior variances of the parameters for the "Cycling Model" after 6864 samples, obtained by inversely modeling the synthetic cycling data with different applied noise levels. The variances are expressed in a logarithmic scale, i.e., the standard deviation measures the width of the distribution in e-folds. This means the upper (lower) bounds of the $k\sigma$-interval of parameter $P$ are given by $P_{\pm} = P_\textrm{mean} \textrm{exp}(\pm k\sigma_P)$. Therefore, comparing the variances shows the relative uncertainty in the specific parameter, which varies depending on the noise level. The prior belief of the parameters is equal for the different noise levels and symmetric around the true parameter values.}
\label{tab.Cov_NoiseLevel}
\end{table}

\begin{figure}[]
	\centering
	\makebox[0pt]{%
    \includegraphics[width=0.43\paperwidth]{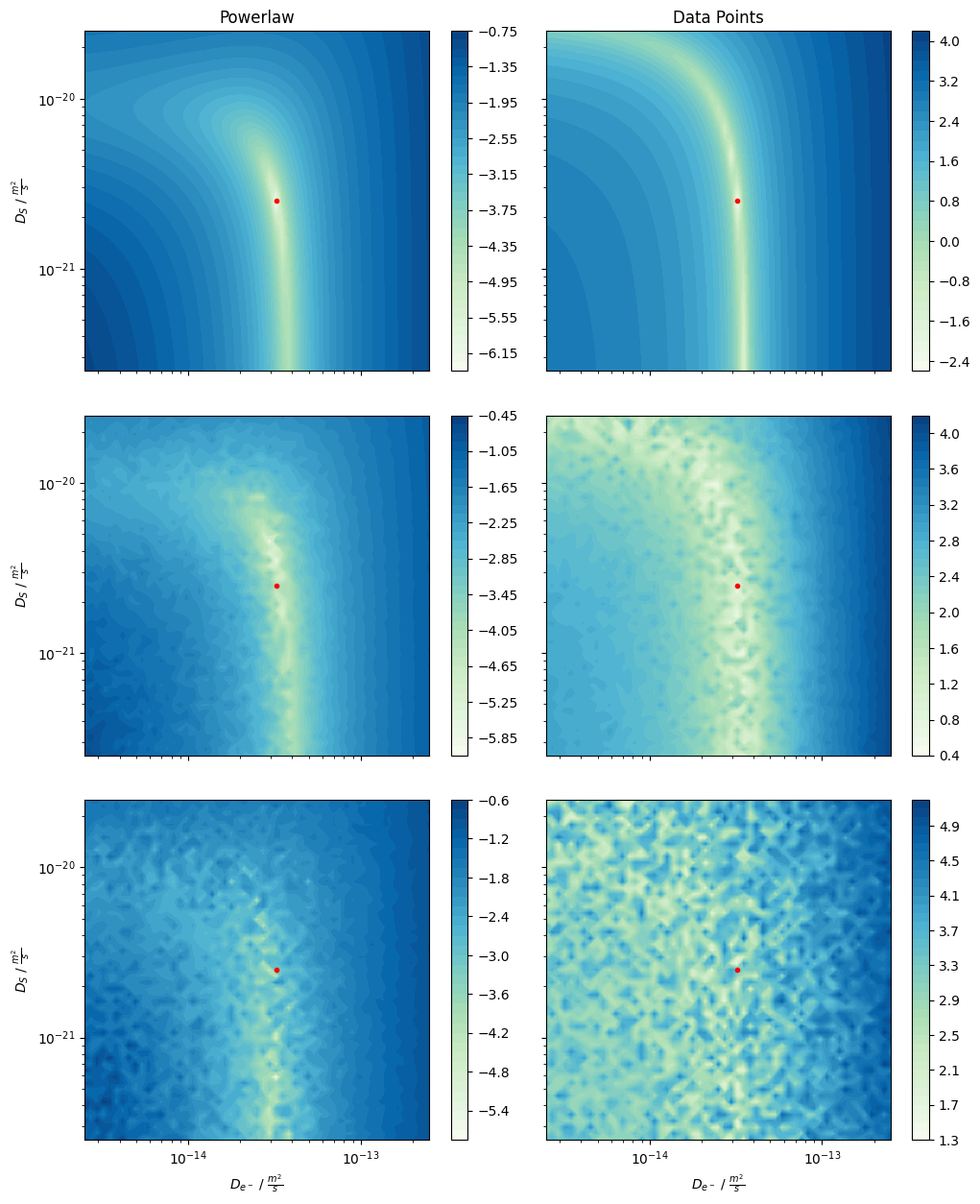}}
	\caption{Landscape of a two-dimensional parameter space ($D_\mathrm{e^-}$-$D_\mathrm{S}$) for different feature choices (columns) at various noise levels (top row: $\sigma_\textrm{noise}^2 = 0 \textrm{Ah}^2$, middle row: $\sigma_\textrm{noise}^2 = 8 \cdot 10^{-6} \textrm{Ah}^2$, bottom row: $\sigma_\textrm{noise}^2 = 8 \cdot 10^{-5} \textrm{Ah}^2$). The red dots mark the true parameter configuration. In the left column, one power law through the capacity loss of 500 cycles is considered one feature ($f = [\alpha, \beta]^{\mathrm{T}}$). In the right column, the data points of the 500 cycles without transformation are considered as one feature ($f = [\textrm{CL}(t_\textrm{Cycle 1}), ..., \textrm{CL}(t_\textrm{Cycle 500})]^{\mathrm{T}}$). The color indicates the value of the loss function, which is given as the relative distance of the feature applied to the simulated and experimental data: $L = \textrm{log}(\vert \vert \frac{f_i(y_\mathrm{sim}(\theta))}{f_i(y_\mathrm{exp})} -1 \vert \vert)$.}
	\label{fig.Cycl_PSpace_PL_Euc_Full_NoiseComp}
\end{figure}

A central aspect of parameterization is how to measure the distance between data curves, e.g., experimental data and simulated data. The chosen transformations in the features act as such metrics because they assign each parameter combination a certain distance to the optimum (experimental data). Ideally, this metric can identify the correct parameter combination independently of the noise level. Therefore, we investigate a visualization of these metrics. Figure \ref{fig.Cycl_PSpace_PL_Euc_Full_NoiseComp} shows a two-dimensional cross-section of the landscape in parameter space for different features (columns) and noise levels (rows) at the correct value for $\kappa_\mathrm{{Li}^+,SEI}$. The red dots indicate the true parameters. With vanishing noise, the landscape of the power law feature (upper left panel) shows a global minimum for the true parameters. Considering the data points (upper right panel) after each cycle as a feature (i.e. $f = [\textrm{CL}(t_\textrm{Cycle 1}), ..., \textrm{CL}(t_\textrm{Cycle 500})]^{\mathrm{T}}$), only a valley of minima can be identified rather than a global minimum. For higher noise levels, the data point feature shows a vast and smeared-out valley of possible parameter combinations and is therefore considered a bad feature choice. In contrast, the power law feature maintains a narrow area around the correct parameters even for higher noise levels. Choosing certain features can be interpreted as inputting the visualized landscapes into the optimization algorithm. Therefore, a reasonable choice of features, their combinations, and noise resilience is crucial for correct parameterization, uncertainty quantification, and correlations. Furthermore, they directly impact the convergence speed of the algorithm and the needed number of simulated samples. For a more detailed discussion of features, see SI \ref{sec:Feat_Choice}.


\section{Conclusion}\label{sec:Conclusion}

In this work, we investigate physics-based degradation models implemented in a full battery cell model with Bayesian ML algorithms. The resulting parameterization and uncertainty quantification have to be correct and precise within the order of thousand simulated samples. Moreover, correct ML-based identification of the prevailing degradation model has to be achieved.

For synthetic storage data, the EP-BOLFI algorithm can indeed find the correct parameterization and a reasonable corresponding uncertainty within a few samples. Further, valuable correlation values between the parameters are obtainable for the posterior distribution and self-selected features. We show that these correlations correctly reflect the physical interplay between different processes and can show relevant processes in the chosen features.

By analyzing real storage data, we show that SEI growth by electron conduction and electron diffusion can describe a similar trend in capacity loss with increasing storage SoC. EP-BOLFI can parameterize these models with reasonable uncertainty. To identify the best model for the experimental data, we investigate a Bayesian model selection criterion. This method consistently identifies the correct model and favors electron diffusion as the relevant SEI growth mechanism during battery storage for the investigated experimental data.

Further, we show that inverse modeling of noisy cycling data is possible for a significantly biased and uncertain prior belief. We identify the correct contribution and interplay of multiple SEI growth mechanisms. We visualize the importance of a suitable choice of features for convergence speed, obtained uncertainty, noise stability, and correct parameter identifiability. Further, we find that the resulting uncertainty correctly depends on the noise level in the data.

To conclude this work, using sample-efficient Bayesian algorithms enables the inverse modeling of real physics-based models within acceptable computation time. The results contain correct parameterization with physically reliable uncertainty quantification and summary statistics. We show these methods identify electron diffusion responsible for the SoC-dependent capacity loss during battery storage. However, for further confirmation, it is of major importance to analyze all-encompassing experimental data measured at the most stable conditions and cell chemistries. Only then can the background of the SoC-independent capacity loss, the influence of check-ups, and the actual SEI growth mechanisms be identified.

To obtain the needed profound physical understanding of battery processes, we propose to inversely analyze developed models with the presented methods to identify the correct physical process descriptions.

\vspace*{0.5cm}

\begin{acknowledgments}
This work was funded by the European Union's Horizon 2020 research and innovation programme under grant agreement No. 957189 and No. 101103997. The authors acknowledge support by the state of Baden-Württemberg through bwHPC and the German Research Foundation (DFG) through grant No. INST 40/575-1 FUGG (JUSTUS 2 cluster). Further, the authors acknowledge support by the Helmholtz Association through grant No. KW-BASF-6 (Initiative and Networking Fund as part of the funding measure "ZeDaBase-Batteriezelldatenbank").
\end{acknowledgments}

\bibliography{MLBOPaperMendeley}

\appendix

\section{Penalty in BASQ}\label{sec:BASQ_Penalty}

The likelihood $LL$ used in this work originates from the implementation in BASQ \cite{Adachi2023}

\begin{equation}\label{eq:BASQ_LL}
\begin{split}
    LL = &- \sum\limits_{j=1}^n \mathrm{log}(2 \pi R) \\
    &- \frac{\vert \vert y_\textrm{exp} - y_\textrm{sim} \vert \vert +\lambda \textrm{Tr}(\sum_\textrm{Prior})\sum\limits_{i=1} (\frac{\theta_i}{\theta_{i,0}})^2}{2 R},
\end{split}
\end{equation} 

where $n$ is the number of data points, $R$ describes the noise in the data ($R = \mathrm{exp}(-\mathrm{exp}(\sigma_\textrm{noise}^2))$), $\lambda$ is the parameter to adjust the weighting of the penalty, $\sum_\textrm{Prior}$ is the covariance matrix of the prior, $\theta_i$ is the value of parameter $i$ and $\theta_{i,0}$ is the mean value of the prior of parameter $i$. For this work, we introduced the penalty term, which consists of the trace of the prior covariance matrix and parameter values themselves. Thereby, more available parameters are penalized by the trace of the covariance matrix, and the optimizing scheme is motivated to lower the parameter values as much as possible. In the used case, this is equivalent to Occam's razor since lowering the parameters as much as possible equals leaving out the corresponding effect. This penalty has to be calibrated by adjusting $\lambda$ to not minimize the penalty alone but to find the optimal parameter configuration and still penalize more complex models.

\section{SPMe equations and conditions}\label{sec:SPMe_Tables}

In Table \ref{tab.SPMe_nodim}, the non-dimensional form of the governing equations of the SPMe model, as well as the equations for the terminal cell voltage, are listed. Note that the overbar indicates an electrode-averaged quantity. Boundary and initial conditions are needed to complete and solve the set of differential equations. These are shown in Tab. \ref{tab.SPMe_nodim_BConditions} and Tab. \ref{tab.SPMe_nodim_IConditions}, respectively.

\begin{table*}[t!]
\begin{center}
\caption{Governing equations of SPMe battery model}
\label{tab.SPMe_nodim}
\hrule
\begin{subequations}
\begin{align}
    &\text{particle radial diffusion} &\mathcal{C}_\mathrm{k} \partial_\mathrm{t} c_\mathrm{s,k}^0 &= - \frac{1}{r_\mathrm{k}^2} \partial_\mathrm{r_\mathrm{k}}(r_\mathrm{k}^2 \partial_\mathrm{r_\mathrm{k}}c_\mathrm{s,k}^0) \label{eq:nodim-ie}&& \\
    &\text{electrolyte cation molar flux} & N_\mathrm{e,k}^1 &= -\epsilon_\mathrm{k}^\mathrm{b}D_\mathrm{e}(1) \partial_\mathrm{x} c_\mathrm{e,k}^1 + \begin{cases}
                                             \frac{xt^+I}{\gamma_\mathrm{e}L_\mathrm{n}},& \mathrm{k = n}\\
                                             \frac{t^+I}{\gamma_\mathrm{e}},& \mathrm{k = s}\\
                                             \frac{(1-x)t^+I}{\gamma_\mathrm{e}L_\mathrm{p}},& \mathrm{k = p}
                                         \end{cases} \label{eq:nodim-ie}&& \\
    &\text{electrolyte cation diffusion} & \mathcal{C}_\mathrm{e} \epsilon_\mathrm{k} \gamma_\mathrm{e} \partial_\mathrm{t} c_\mathrm{e,k}^1 &= - \gamma_\mathrm{e} \partial_\mathrm{x} N_\mathrm{e,k}^1 + \begin{cases}
                                             \frac{I}{L_\mathrm{n}},& \mathrm{k = n}\\
                                             0,              & \mathrm{k = s}\\
                                             -\frac{I}{L_\mathrm{p}},& \mathrm{k = p}
                                         \end{cases} \label{eq:nodim-ie}&& \\
    &\text{terminal voltage} &V &= \overline{U}_\mathrm{eq} + \overline{\eta}_\mathrm{r} + \overline{\eta}_\mathrm{c} + \overline{\Delta \Phi}_\mathrm{Elec} + \overline{\Delta \Phi}_\mathrm{Solid} \label{eq:nodim-ie}&& \\
    &\text{equilibrium voltage} &\overline{U}_\mathrm{eq} &= U_\mathrm{p}(c_\mathrm{s,p}^0 \vert_{r_\mathrm{p}=1}) - U_\mathrm{n}(c_\mathrm{s,n}^0 \vert_{r_\mathrm{n}=1}) \label{eq:nodim-ie}&& \\
    &\text{reaction overpotential} &\overline{\eta}_\mathrm{r} &= - 2\mathrm{sinh}^{-1}\left(\frac{I}{\overline{j}_\mathrm{0,p} L_\mathrm{p}}\right) - 2\mathrm{sinh}^{-1}\left(\frac{I}{\overline{j}_\mathrm{0,n} L_\mathrm{n}}\right)\label{eq:nodim-ie}&& \\
    &\text{concentration overpotential} &\overline{\eta}_\mathrm{c} &= 2\mathcal{C}_\mathrm{e}(1-t^+)(\overline{c}_\mathrm{e,p}^1 - \overline{c}_\mathrm{e,n}^1) \label{eq:nodim-ie}&& \\
    &\text{negative electrode exchange current densities} &\overline{j}_\mathrm{0,n} &= - \frac{1}{L_\mathrm{n}} \int_{0}^{L_\mathrm{n}} \frac{\gamma_\mathrm{n}}{\mathcal{C}_\mathrm{r,n}}(c_\mathrm{s,n}^0)^{1/2}(1-c_\mathrm{s,n}^0)^{1/2}(1+\mathcal{C}_\mathrm{e}c_\mathrm{s,n}^1)^{1/2}\mathrm{d}x \label{eq:nodim-ie}&& \\
    &\text{positive electrode exchange current densities} &\overline{j}_\mathrm{0,p} &= - \frac{1}{L_\mathrm{p}} \int_{1-L_\mathrm{p}}^{L_\mathrm{p}} \frac{\gamma_\mathrm{p}}{\mathcal{C}_\mathrm{r,p}}(c_\mathrm{s,p}^0)^{1/2}(1-c_\mathrm{s,p}^0)^{1/2}(1+\mathcal{C}_\mathrm{e}c_\mathrm{s,p}^1)^{1/2}\mathrm{d}x \label{eq:nodim-ie}&& \\
    &\text{electrolyte ohmic losses} &\overline{\Delta \Phi}_\mathrm{Elec} &= - \frac{I}{\hat{\kappa}_\mathrm{e}\kappa_\mathrm{e}(1)} \left(\frac{L_\mathrm{n}}{3\epsilon_\mathrm{n}^\mathrm{b}}+\frac{L_\mathrm{s}}{\epsilon_\mathrm{s}^\mathrm{b}}+\frac{L_\mathrm{p}}{3\epsilon_\mathrm{p}^\mathrm{b}}\right) \label{eq:nodim-ie}&& \\
    &\text{solid phase ohmic losses} &\overline{\Delta \Phi}_\mathrm{Solid} &= -\frac{I}{3}\left(\frac{L_\mathrm{p}}{\sigma_\mathrm{p}}-\frac{L_\mathrm{n}}{\sigma_\mathrm{n}}\right) \label{eq:nodim-eta}&&
\end{align}
\end{subequations}
\hrule
\end{center}
\end{table*}

\begin{table*}[t!]
\begin{center}
\caption{Boundary conditions}
\label{tab.SPMe_nodim_BConditions}
\hrule
\begin{subequations}
\begin{align}
    &\text{particle} &\partial_\mathrm{r_\mathrm{k}} c_\mathrm{s,k}^0 \vert_{r_k = 0} &= 0, \:  -\frac{a_k \gamma_k}{\mathcal{C}_\mathrm{k}} \partial_\mathrm{r_\mathrm{k}}=
    \begin{cases}
                                            \frac{I}{L_\mathrm{n}},& \mathrm{k = n}\\
                                             -\frac{I}{L_\mathrm{p}},& \mathrm{k = p}\\
    \end{cases}\label{eq:nodim-ie}&& \\
    &\text{molar flux} & N_\mathrm{e,n}^1 \vert_{x=0} &= 0, \: N_\mathrm{e,p}^1 \vert_{x=1} = 0, \: N_\mathrm{e,n}^1 \vert_{x=L_\mathrm{n}} = N_\mathrm{e,s}^1 \vert_{x=L_\mathrm{n}}, \: N_\mathrm{e,s}^1 \vert_{x=1-L_\mathrm{p}} = N_\mathrm{e,p}^1 \vert_{x=1- L_\mathrm{p}}, \label{eq:nodim-ie} && \\
    &\text{concentration} &c_\mathrm{e,n}^1 \vert_{x=L_\mathrm{n}} &= c_\mathrm{e,s}^1 \vert_{x=L_\mathrm{n}}, \: c_\mathrm{e,s}^1 \vert_{x=1-L_\mathrm{p}} = c_\mathrm{e,p}^1 \vert_{x=1- L_\mathrm{p}} \label{eq:nodim-eta}&&
\end{align}
\end{subequations}
\hrule
\end{center}
\end{table*}

\begin{table}[t!]
\begin{center}
\caption{Initial conditions}
\label{tab.SPMe_nodim_IConditions}
\hrule
\begin{subequations}
\begin{align}
    c_\mathrm{s,k}^0(r_\mathrm{k},0) &=  c_\mathrm{k,0}\label{eq:nodim-ie}&& \\
    c_\mathrm{e,k}^1(x,0) &= 0 \label{eq:nodim-eta}&&
\end{align}
\end{subequations}
\hrule
\end{center}
\end{table}

\section{Temporal evolution of the SEI thickness}\label{sec:SEI_thickness}

By integrating eq. (\ref{eq:dL_SEIdt}) with the single current densities over time while keeping everything despite $L_\mathrm{SEI}$ constant, one can derive with the initial condition ($L_\mathrm{SEI}(t=0) = L_0$) a simplified evolution of the SEI thicknesses,

\begin{subequations}
    \begin{align}
        &L_\mathrm{SEI, ED}  &&= \sqrt{2V_\mathrm{SEI}c_\mathrm{e^-}D_\mathrm{e^-}\mathrm{exp}\left(-\tilde{\eta}_\mathrm{SEI}\right)t + L_0^2} \label{eq:L_ED}, \\
        &L_\mathrm{SEI, EC}  &&= \sqrt{2\frac{V_\mathrm{SEI}}{F}\kappa(\Phi_0 - U(\mathrm{SoC}))t + L_0^2} \label{eq:L_EC},\\
        &L_\mathrm{SEI, SD}  &&= \sqrt{2V_\mathrm{SEI}c_\mathrm{S}D_\mathrm{S}t + L_0^2} \label{eq:L_SD},\\
        &L_\mathrm{SEI, EM}  &&= \frac{V_\mathrm{SEI} c_\mathrm{e^-} D_\mathrm{e^-} F j_\mathrm{int}}{2RT\kappa_\mathrm{{Li}^+,SEI}}\mathrm{exp}\left(-\tilde{\eta}_\mathrm{SEI}\right)t + L_0 \label{eq:L_EM}.
    \end{align}
\end{subequations}
All storage mechanisms (eq. (\ref{eq:L_ED}) - eq. (\ref{eq:L_SD})) can produce the same SEI growth behavior with the square root of time. From this perspective, the proposed storage mechanisms are not distinguishable. However, these mechanisms strongly differ in their dependency on the SoC. Where solvent diffusion is totally SoC-independent, electron conduction shows a slight SoC dependency, which can be varied by the chosen value for $\Phi_0$. In contrast, electron diffusion predicts a strong exponential SoC-dependency through $\tilde{\eta}_\mathrm{SEI}$.

\section{Feature-specific correlations}\label{sec:Feat_Corr}

In the algorithm's structure, the feature-specific summary statistics become calculated backwards (by using eq. (\ref{eq.BayesTheroem})) from the reduced posterior distribution, given the prior. For example, if a particular feature is analyzed, then the likelihood is estimated by a surrogate first. In the second step, the posterior is calculated by incorporating the prior belief. Once this posterior is reduced to a normal distribution, the likelihood can be calculated finally in the form of a normal distribution. The corresponding covariance matrix for the simulated feature is then used to calculate the feature-specific correlations. However, in rare cases, this structure can cause meaningless values, as one can see by the following calculation. Assuming the prior and likelihood distribution to follow a normal distribution, i.e., $P(\theta) \sim \mathcal{N}(\mu, \sigma^2)$ and $P(y\mid \theta) \sim \mathcal{N}(\nu, \tau^2)$, one can show then that the posterior distribution also follows a normal distribution

\begin{equation}
    \mathcal{N}(\rho^2(\frac{\mu}{\sigma^2}+\frac{\nu}{\tau^2}), \rho^2) \sim \mathcal{N}(\nu, \tau^2)\cdot \mathcal{N}(\mu, \sigma^2),
\end{equation}
where $\rho^2 = \left(\frac{1}{\frac{1}{\sigma^2}+\frac{1}{\tau^2}}\right)$. If the posterior becomes more uncertain about a parameter distribution during the analysis of the feature likelihood, i.e., $\rho^2>\sigma^2$, this can only be realized with $\tau^2<0$. Hence, no valuable summary statistics are available for this feature. 

\begin{figure}[]
	\centering
	\makebox[0pt]{%
    \includegraphics[width=0.43\paperwidth]{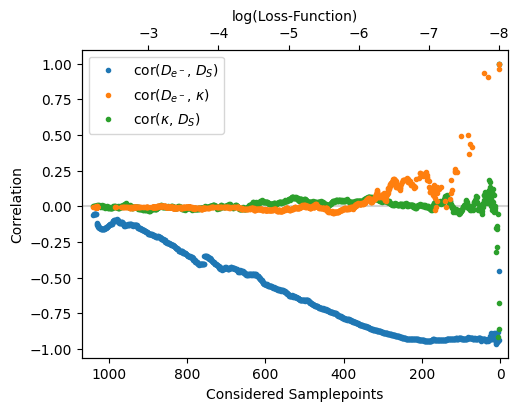}}
	\caption{Feature-specific correlations for the low-SoC-feature in the inverse modeling of the synthetic data with the "Overfitted Model" ($\Phi_0 =0.145$V) and a total of 1040 samples. Each dot indicates the correlation value between two parameters when considering a certain number of samples. On the left, all 1040 drawn samples are considered. Moving to the right, the worst samples with the highest discrepancy to the synthetic data become left out. On the right, only the few best samples are considered.}
	\label{fig.SynthStor_FeatCorr}
\end{figure}

There are multiple ways still to get meaningful correlation values for a specific feature. One possibility is to do the inverse modeling for the feature of interest only, using the already obtained parameterization from the entire run. Then, the posterior contains only the information from the specific feature and is always well-posed. It is important to note that the prior itself influences the result depending on the precision achieved in the feature. Therefore, enough samples with good precision have to be simulated to diminish the impact of the prior. Since EP-BOLFI uses the information carried by the best samples (see threshold $\epsilon$ in eq. (\ref{eq:LL_threshold})) to deduce the probability distribution, it is further crucial that enough samples achieve equal good precision as illustrated in Fig. \ref{fig.SynthStor_FeatCorr}. The data points were obtained by inversely model the synthetic storage data with the "Overfitted Model" (with 1040 samples) using only the low-SoC feature. The colored dots refer to the correlation values between the different model parameters in the low-SoC feature, depending on the number of considered samples. From left to right, the number of considered samples to calculate the correlation values decreases (lower x-axis), whereas the precision of the considered samples increases (upper x-axis). On the left side, all correlation values are close to zero, as all randomly drawn samples are considered. However, moving along the x-axis, only the samples that accurately describe the feature are considered, and real correlations emerge. Since electron conduction (with $\Phi_0 =0.145$V) doesn't contribute to the low-SoC feature, the resulting correlation values with $\kappa$ should be around zero, which is the case for most data points. This can change drastically if the number of considered samples is too low since the values for $\kappa$ are drawn randomly and might show a positive or negative correlation by chance only (see orange and green dots for 0-200 sample points). Therefore, the second possibility to obtain reasonable correlations for a specific feature is to create this kind of correlation figures from the samples simulated in EP-BOLFI for the feature of interest.

\section{Feature-choice}\label{sec:Feat_Choice}

\begin{figure*}[]
	\centering
	\makebox[0pt]{%
    \includegraphics[width=0.85\paperwidth]{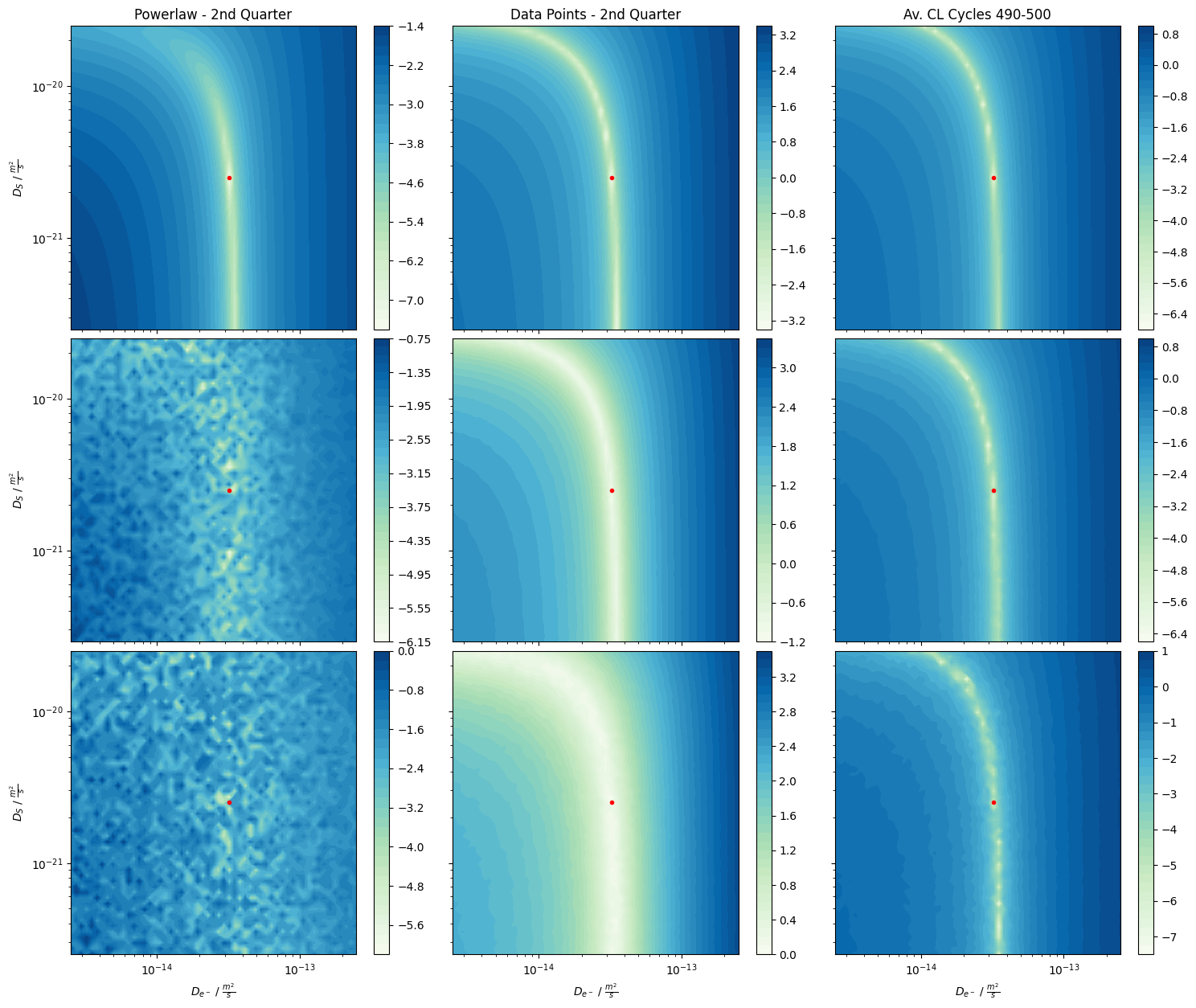}}
	\caption{Landscape of a two-dimensional parameter space for different feature choices (columns) at various noise levels (top row: $\sigma_\textrm{noise}^2 = 0 \textrm{Ah}^2$, middle row: $\sigma_\textrm{noise}^2 = 8 \cdot 10^{-6} \textrm{Ah}^2$, bottom row: $\sigma_\textrm{noise}^2 = 8 \cdot 10^{-5} \textrm{Ah}^2$). The red dots mark the true parameter configuration. In the left column, one power law through the capacity loss of a subset of cycles (cycles 126 to 250) is considered one feature ($f = [\alpha, \beta]^{\mathrm{T}}$). This is in contrast to the left column of Fig. \ref{fig.Cycl_PSpace_PL_Euc_Full_NoiseComp}, where the powerlaw is applied to all data points. In the middle column, the data points of cycles 126 to 250 without transformation are considered one feature ($f = [\textrm{CL}(t_\textrm{Cycle 126}), ..., \textrm{CL}(t_\textrm{Cycle 250})]^{\mathrm{T}}$). In the right column, the average of the data points of the last ten cycles is considered as one feature ($f = [\overline{\textrm{CL}}(t_\mathrm{av})]$). The color indicates the value of the loss function, which is given as the relative distance of the feature applied to the simulated and experimental data: $L = \textrm{log}(\vert \vert \frac{f_i(y_\mathrm{sim}(\theta))}{f_i(y_\mathrm{exp})} -1 \vert \vert)$. This emphasizes that suitable feature-transformations and their noise resilience strongly depend on the underlying problem.}
	\label{fig.Cycl_PSpace_PL_Euc_Full_Av_NoiseComp}
\end{figure*}

As the feature determines which metric is applied to measure the distance between the simulated and the experimental data, it directly influences the assigned likelihood for each drawn parameter configuration. The landscape of the loss function for the possible parameter combination visualizes this information. The existence of global minima refers to a unique optimal parameterization, whereas valleys indicate multiple optimal parameter combinations. The shape of such valleys contains information about the expected correlation values, which depend on the location in the parameter space. The width of the valleys represents the uncertainty in the parameterization. 
Different features capture different aspects of the data, e.g., power laws refer more to the actual shape of the curve in data space. In contrast, data points themselves, as a feature, try to minimize only the overall distance to the experimental data undisturbed by the actual shape. This can be seen in Fig. \ref{fig.Cycl_PSpace_PL_Euc_Full_Av_NoiseComp} as the feature for the averaged capacity loss of the last ten cycles shows a very sharp valley in the $D_\mathrm{e^-}$-$D_\mathrm{S}$ plane. The main objective of this feature is to capture a single value of the CL at the end of the experiment. Therefore, it is not interested in the actual trajectory describing this capacity loss. Then, there are multiple solutions to tune electron diffusion, solvent diffusion, or electron migration to capture this point. However, not all possible solutions present the right trajectory, as the power law feature indicates only a subarea as optimal. The intersection of the minima of different features will yield the final parameterization. Suitable features vary enormously depending on the underlying optimization problem or differ even for one specific problem. For example, looking at power laws for small parts of noisy data only performs worse than considering the data points in the same parts (see lower panels of Fig. \ref{fig.Cycl_PSpace_PL_Euc_Full_Av_NoiseComp}). In contrast, a power law through all data is less sensitive to noise than considering all data points (see Fig. \ref{fig.Cycl_PSpace_PL_Euc_Full_NoiseComp}). Combining a smart choice of features, which captures different aspects of the data, enables a fast, correct and certain parameterization.

\end{document}